# FLARECAST: an I4.0 technology for space weather using satellite data

Michele Piana
Dipartimento di Matematica
Università di Genova
Genova, Italy

Anna Maria Massone
CNR - SPIN, Genova
Genova, Italy

Federico Benvenuto
Dipartimento di Matematica
Università di Genova
Genova, Italy

Cristina Campi
Dipartimento di Medicina
Università di Padova
Padova, Italy

*Abstract*—'Flare Likelihood and Region Eruption Forecasting (FLARECAST)' is a Horizon 2020 project, which realized a technological platform for machine learning algorithms, with the objective of providing the space weather community with a prediction service for solar flares. This paper describes the FLARECAST service and shows how the methods implemented in the platform allow both flare prediction and a quantitative assessment of how the information contained in the space data utilized in the analysis may impact the forecasting process.

*Keywords—space weather; solar flares; flare prediction; machine learning*

## I. INTRODUCTION

Space weather [1] is a rather modern term, which refers to the impact that the dynamical conditions of the solar atmosphere may have on the heliosphere and, more specifically, on the geosphere. The main drivers of adverse space weather are solar flares and Coronal Mass Ejections (CMEs), which, in turn, represent the strongest energetic events in the solar system, influencing a panorama of physical systems from the solar surface, through the inner heliosphere and onwards into geo-space and down to Earth. This possible stream of events, from time to time significantly affect the performance of Earth- and space-based technological systems, mainly terrestrial communications links, power grids, and satellite operations.

Space weather can be rather intriguingly interpreted as an 'Industry 4.0 (I4.0)' issue. In fact, twenty-eight space missions coordinated by the most important space agencies worldwide are currently under either primary or extended operation status with scientific objectives in several ways related to space weather topics (eight more missions will be launched in the next five years). This impressive fleet of space satellites, travelling both terrestrial and more complicated orbits, may be viewed as a gigantic 'Internet of Things (IoT)' network of sensors systematically monitoring the space weather conditions and their impact on on-Earth technologies. Further, telescopes on-board this fleet continuously collect electromagnetic and particle measurements related to solar flares, CMEs and their by-products and this huge amount of observations must be transferred, stored, and handled according to hardware/software solutions typical of standard 'big-data' science. Eventually, extraction of meaningful information from these data can be performed just by means of sophisticated artificial intelligence tools that allow the identification of clusters of signals via pattern recognition algorithms, the reconstruction of parameters with specific physical significance via regularization methods for inverse problems, the prediction of the dynamics of such parameters via algorithms formulated within machine learning frameworks.

One of the most systematic attempt to deal with space weather data according to this 'I4.0' perspective is probably represented by the Horizon 2020 Research and Innovation Action 'Flare Likelihood and Region Eruption Forecasting (FLARECAST)'. This project effort within the 'PROTEC-1-2014 Space Weather' call had the objective to realize a flare prediction system based on automatically extracted physical properties of active regions coupled with machine-learning-based flare prediction methods and validated using the most appropriate forecast verification measures. FLARECAST technological platform imports data from an 'ad hoc' set up remote archive of magnetograms recorded by the 'Helioseismic and Magnetic Imager' in the 'Solar Dynamics Observatory' (SDO/HMI) [2], automatically extracts features from the loaded data, models such features within the framework of machine learning methods and provides a final verification of results. The implementation of the FLARECAST software prototype is conceived in a highly modular fashion and is therefore an extendable architecture, whereby further algorithms can be added to the platform and systematically validated.

The objective of the present paper is two-fold. First, we will provide a quick overview of the FLARECAST service, with specific emphasis of the machine learning methods that currently represent its state-of-the-art. Second, we will show how the FLARECAST service can be used for flare forecasting and feature ranking, even more importantly, to identify which extracted features mostly impact the forecasting effectiveness.

The plan of the paper is as follows. In Section 2 we will describe the current status of the data analysis service that FLARECAST may provide to the space weather community. Section 3 discusses how the service can be used to realize flare prediction and feature ranking. Our conclusions will be offered in Section 4.

## II. THE FLARECAST ARCHITECTURE

Figure 1 shows the current modular architecture of the FLARECAST infrastructure. This architecture follows the four core steps of the FLARECAST processing workflow, i.e.:

1. Step 1 imports data from the archive of SDO/HMI (although other remote archives can be used, such as the one at the NOAA Space Weather Prediction Center).
2. Step 2 extracts features from the loaded data
3. Step 3 models the data features by means of machine learning methods.
4. Step 4 validates the results from the previous step.

The computational core of this pipeline is of course represented by Step 3. The machine learning methods currently implementing this step can be divided into the two families of unsupervised and supervised methods, whereby unsupervised approaches realize an automatic clustering of the observations according to homogeneity properties and without the use of any historical dataset, while supervised approaches exploit a labeled historical dataset to train the prediction when unlabeled data arrive for the analysis. For both families, we implemented standard approaches that have been already utilized for flare forecasting purposes; advanced machine learning methods that have been developed within FLARECAST in order to fulfill objectives specific for this kind of application (for example, ad hoc designed loss functions in regularization networks, penalty terms tailored to feature selection requirements, more automatic and more general approaches to unsupervised classification); and, finally, innovative approaches that have been formulated within the project framework and that are applied here for the first time against a real-world use-case. Specifically, the unsupervised methods currently at disposal of the FLARECAST users are: K-means [3], Fuzzy C-means [4], possibilistic C-means [5], and simulated annealing [6]. The supervised methods that can be used in the present version of the platform are: LASSO [7], hybrid LASSO [8], elastic net [9], hybrid logit [8], random forest [10], a multi-layer perceptron [11], a recurrent neural network [12], a support vector machine with different kinds of kernels [13], the Garson method [14], and the Olden method [15]. The overall service is at disposal for immediate download to local machines while the on-line service is currently under construction

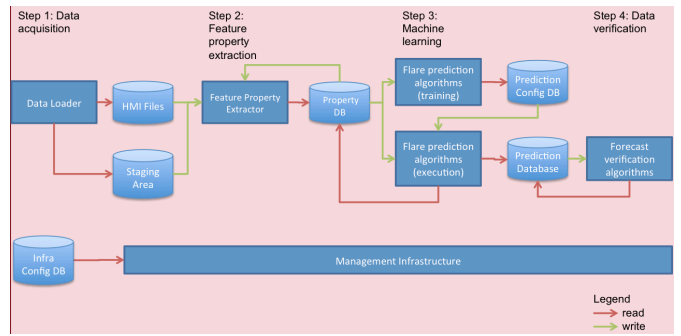

Figure 1: the FLARECAST platform

## III. AN EXAMPLE OF FLARE FORECASTING WITH FLARECAST

Solar flares [16] are explosive events in which huge amounts of energy, previously stored in coronal magnetic field configurations, are released in a time scale between 10s and 1000s, thus triggering secondary effects like the emission of electromagnetic radiation at all wavelengths, coronal materials, and energetic particles. Solar active regions (ARs) are, undeniably, the major hosts of flares and therefore any flare prediction process must rely on the availability of experimental information on them and, in particular, on their magnetic properties. SDO/HMI provides full disk magnetograms with unprecedented spatial and temporal resolution and from them it is possible to extract a high number of properties of the ARs presented in each image.

We now describe an example of how FLARECAST is able to exploit such properties in order to realize a complex flare prediction task. We considered a set of point-in-time SDO/HMI images with a temporal cadence of 24 hours in the time range between September, 14 2012 and April, 30 2016. The feature extraction algorithms implemented in FLARECAST allow the user to extract up to 171 features (i.e., geometrical and physical properties) associated to each AR present in the point-in-time image (this is, possibly, the richest dataset possibly at disposal for this kind of analysis in the scientific community). This implies that, for the considered time range, we can construct 4442 set of 171-dimension feature vectors (one has to consider that a single AR may last for more than one HMI image). Since the machine learning analysis we want to perform in this example is of supervised kind, we construct a training set by randomly extracting 66% ARs from the overall set of ARs and label the feature vectors associated to each extracted AR by annotating whether a flare with an intensity sufficiently high occurred in the next 24 hours (solar flares are classified according to their intensity, i.e. to the amount of energy they release: from low to high energy we have flares of class C, class M, and class X, respectively; here we consider flares of class C or higher). The set of the remaining feature vectors is not labeled and is used as test set for the experiment. Given this training set we want to analyze the test set in order to solve two problems, i.e. to

- Predict whether an at least C flare occurred in the next 24 hours.

- Determine which features among the 171 ones mostly impacted such prediction (this means to compute the weights with which the features contributed to the prediction and rank them).

We performed this analysis by means of the hybrid LASSO algorithm [8], although other methods performing feature ranking could be utilized. In order to describe this two-step approach, let us denote with $X$ is the $N \times F$ matrix in which $N = 4442$ is the number of available feature vectors of dimension $F = 171$. Then, $y$ is the $N \times 1$ vector made of the binary labels in the training set and, finally, $\beta$ is the $F \times 1$ vector made of feature weights. Therefore, in this setup, $X$ represents the training set, $y$ is the input data and $\beta$ is the unknown. In fact, the first, LASSO step of the method computes

$$\hat{\beta} = argmin_\beta(\|y - X\beta\|_2^2 + \lambda\|\beta\|_1)$$

while the second step applies an unsupervised clustering algorithm to $\hat{y} = X\hat{\beta}$ in order to partition the components of $\hat{y}$ into two classes, which corresponds to determine a data-adaptive threshold: when a new feature vector $x$ arrives, the algorithm computes $x^t\hat{\beta}$ and assigns it to the closest class; in this way the prediction is made, while each component of $\hat{\beta}$ provides a quantitative assessment of the impact of the corresponding feature. Having at disposal a test set, we may perform a quantitative assessment of the prediction method. Indeed, several skill scores exist that estimate the ability of forecasting methods to realize a binary prediction [17]. As typical examples and denoting with $TP$ the number of true positives, $FP$ the number of false positives, $TN$ the number of true negatives, and $FN$ the number of false negatives, possible skill scores are: the True Skill Statistics

$$TSS = \frac{TP}{TP + FN} - \frac{FP}{FP + TN}$$

the Heidke Skill Score

$$HSS = \frac{2(TP \cdot TN - FN \cdot FP)}{(TP + FN) \cdot (FN + TN) + (TP + FP) \cdot (FP + TN)}$$

the Accuracy

$$ACC = \frac{TP + TN}{TP + TN + FP + FN}$$

and the False Alarm Ratio

$$FAR = \frac{FP}{TP + FP}$$

When applied to the test set defined at the beginning of the Section, hybrid LASSO provides the numbers in Table 1 for these skill scores, where the standard deviation is computed by means of 100 random realizations of the training/test sets.

| TSS | HSS | ACC | FAR |
|---|---|---|---|
| $0.51 \pm 0.04$ | $0.54 \pm 0.03$ | $0.83 \pm 0.01$ | $0.29 \pm 0.03$ |

Table 1. Performance of hybrid LASSO measured by means of four standard skill scores.

| feature | averaged rank |
|---|---|
| sharp_kw/hgradbh/total | 3.47 |
| r_value_br_logr | 3.52 |
| wlsg_br/value_int | 3.74 |
| wlsg_blos/value_int | 3.96 |
| flare_index_past | 13.98 |

Table 2. Average rank of the ten features that mostly impact the prediction computed by the LASSO step of hybrid LASSO.

In order to check the behavior of hybrid LASSO while assessing the role of each feature in the prediction process, we have computed $\hat{\beta}$ for the 100 random realization of the training/test sets, determined the rank of each component for each realization and averaged the ranks over the realizations. Table 2 illustrates the result of this procedure for the five features that have the best average rank among the 171 components of $\hat{\beta}$.

IV. CONCLUSIONS

This paper describes some potentialities of the flare forecasting service based on machine learning, realized within the framework of the Horizon 2020 Research and Innovation Action 'Flare Likelihood and Region Eruption Forecasting (FLARECAST)'. In particular, we have overviewed the technology and the methods implemented in the service and showed how a specific machine learning method at disposal in the service is able to provide both binary prediction and feature ranking. The validation of this analysis should now be performed against the notable amount of data available from the SDO/HMI archive. Further, the physical implications of the results may represent potential breakthrough in the interpretation of the mechanisms at the basis of solar flare physics. More in general, the actual reliability of FLARECAST and its utility for space weather can be understood just by means of an extended application against many different datasets and for many different problems involving space data in heliophysics.

ACKNOWLEDGMENT

The authors warmly thank all the members of the FLARECAST consortium, and, in particular, the FLARECAST PI Manolis Georgoulis, for the many helpful discussions that have characterized this common effort.

REFERENCES

[1] B. B. Poppe and K. P. Jorden, Sentinels of the Sun: Forecasting Space Weather. Boulder: Johmson, 2006.
[2] J. Schou, P. H. Scherrer et al, "Design and ground calibration of the Helioseismic and Magnetic Imager (HMI) instrument on the Solar Dynamics Observatory (SDO)", Sol. Phys., vol. 275, pp. 229-259, 2012.

[3] M. R. Aldenberg, Cluster Analysis for Applications, London: Academic, 1973.

[4] J. C. Bezdek, Pattern Recognition with Fuzzy Objective Function Algorithms, New York: Plenum, 1981.

[5] A. M. Massone, L. Studer, and F. Masulli, "Possibilistic clustering approach to trackless ring pattern recognition in RICH counters", Int. J. Appr. Reas., vol. 41, pp. 96-109, 2006.

[6] S. Kirkpatrick, C. D. Gelatt, and M. P. Vecchi, "Optimization by simulated annealing", Science, vol. 220, pp. 671-680, 1983.

[7] R. Tibshirani, "Regression shrinkage and selection via the LASSO", J. Roy. Stat. Soc. B, vol 58, pp. 267-288, 1996.

[8] F. Benvenuto, M. Piana, C. Campi, and A. M. Massone, "A hybrid supervised/unsupervised machine learning approach to solar flare prediction", Astrophys. J., vol. 853, p. 90, 2018.

[9] H. Zou and T. Hastie, "Regularization and variable selection via the Elastic Net", J. Roy. Stat. Soc. B, vol. 67, pp. 301-320, 2005.

[10] L. Breiman, "Random forests", Machine Learning, vol. 45, pp. 5-32, 2001.

[11] D. E. Rumelhart, G. E. Hinton, and R. J. Williams, "Learning representations by back-propagating errors", Nature, vol. 323, pp. 533-536, 1986.

[12] J. L. Elman, "Finding structure in time", Cogn. Sci., vol. 14, p. 179, 1990.

[13] C. Cortes and V. Vapnik, "Support-vector networks", Machine Learning, vol. 20, pp.271-297, 1995.

[14] G. D. Garson, "Interpreting neural-network connection weights", Artif. Intell. Expert, vol. 6, pp. 47-51, 1991.

[15] J. D. Olden, M. K. Joy, and R. G. Death, "An accurate comparison of methods for quantifying variable importance in artificial neural networks using simulated data", Ecol. Model, vol. 178, pp. 389-397, 2004.

[16] E. P. Kontar, J. C. Brown et al, "Deducing electron properties from hard X-ray observations", Space Sci. Rev., vol 159, pag. 301, 2011.

[17] D. S. Bloomfield, P. A. Higgins, R. T. J. McAteer, and P. T. Gallagher, "Toward reliable benchmarking of solar flare forecasting methods", Astrophys. J. Lett., vol. 747, p. L41, 2012.

[18] D.